\documentclass[a4paper, amsfonts, amssymb, amsmath, nofootinbib, reprint, prapplied, aps, showkeys,  twoside,superscriptaddress,noeprint,nolongbibliography]{revtex4-2}
\usepackage[english]{babel}
\usepackage[utf8]{inputenc}
\usepackage[colorinlistoftodos, color=green!40, prependcaption]{todonotes}
\usepackage{multirow}
\usepackage{url}

\usepackage{siunitx}

\usepackage{amsthm}
\usepackage{amsmath}

\usepackage{mathtools}
\usepackage{physics}
\usepackage{xcolor}
\usepackage{float}
\usepackage[left=23mm,right=13mm,top=35mm,columnsep=15pt]{geometry} 
\usepackage{adjustbox}
\usepackage{placeins}
\usepackage[T1]{fontenc}
\usepackage{lipsum}
\usepackage{csquotes}

\usepackage{IEEEtrantools}

\newcommand{\figref}[1]{Figure \ref{#1}}

\renewcommand{\ket}[1]{|#1\rangle}
\renewcommand{\bra}[1]{\langle #1|}

\renewcommand{\ketbra}[2]{|#1\rangle\!\langle#2|}
\newcommand{\expectation}[3]{\langle#1|#2|#3\rangle}

\newcommand{\Proj}[1]{\ensuremath{\ketbra{#1}{#1}}}

\newcommand{\fid}{\bar F}
\newcommand{\E}{\mathcal{E}}

\newcommand{\Id}{\mathbb{I}}

\renewcommand{\Tr}{\mathrm{Tr}}

\usepackage[pdftex, pdftitle={Article}, pdfauthor={Author}]{hyperref}

\usepackage{graphicx}

\newcommand{\hilb}{\mathcal{H}}

\newcommand{\bigeps}{\mathcal E}

\begin{document}
\title{Selective and efficient quantum process tomography for non trace-preserving maps: a superconducting quantum processor implementation}

\author{Q. Pears Stefano}
    \email[Correspondence email address:]{quimeyps@df.uba.ar}
    \affiliation{Departamento de F\'isica, FCEyN, Universidad de Buenos Aires, Buenos Aires, Argentina}
    \affiliation{Consejo Nacional de Investigaciones Cient\'ificas y T\'ecnicas (CONICET), Argentina.}
\author{I. Perito}
    \affiliation{Departamento de F\'isica, FCEyN, Universidad de Buenos Aires, Buenos Aires, Argentina}
\author{L. Reb\'on}
    \affiliation{Departamento de F\'isica, IFLP - CONICET, Universidad Nacional de La Plata, C.C. 67, 1900 La Plata, Argentina}

\begin{abstract}
Alternatively to the full reconstruction of an unknown quantum process, the so-called selective and efficient quantum process tomography (SEQPT) allows estimating, individually and up to the required accuracy, a given element of the matrix that describes such an operation with a polynomial amount of resources. The implementation of this protocol has been carried out with success to characterize the evolution of a quantum system that is well described by a trace preserving quantum map. Here, we deal with a more general type of quantum process that does not preserve the trace of the input quantum state, which naturally arises in the presence of imperfect devices and system-environment interactions, in the context of quantum information science or quantum dynamics control. In that case, we show that with the aid of {\it a priori} information on the losses structure of the quantum channel, the SEQPT reconstruction can be adapted to reconstruct the non-trace-preserving map. We explicitly describe how to implement the reconstruction in an arbitrary Hilbert space of finite dimension $d$. The method is experimentally verified on a superconducting quantum processor of the IBM Quantum services, by  estimating  several non trace-preserving quantum processes in dimensions up to $d=6$. Our results show that it is possible to efficiently reconstruct non trace-preserving processes, with high precision, and with significantly higher fidelity than when the process is assumed to be trace-preserving. 
\end{abstract}


\maketitle

\section{Introduction}\label{sec:intro}

Characterizing the temporal evolution of quantum systems is a crucial task not only used to quantitatively describe naturally occurring processes, but also for certifying the correct functioning of any device that perform quantum information protocols, like those designed for quantum computing and cryptography, among others \cite{Outeiral2020,GoogleAIQuantum2020,Liao2017,Gisin2007,Llewellyn2020}. The different strategies to achieve this task are usually known as \textit{quantum process tomography} (QPT) protocols \cite{Altepeter2003, Wang2007, Mohseni2008}. Given that quantum channels are linear maps, standard QPT schemes based on a linear inversion method \cite{Nielsen}, are conceptually simple but inefficient in practice, as they require an amount of resources that scales exponentially with the size of the system under study. However, protocols that are both selective (they allow to obtain partial information of the channel without the need to completely reconstruct it) and efficient (the number of required measurements increases with the desired precision but does not depend on the size of the system) have been theoretically developed \cite{Bendersky2008,Bendersky2009} and experimentally tested \cite{Schmiegelow2010, Schmiegelow2011,Gaikwad2018,Gaikwad2022}. These protocols make use of particular sets of states known as uniform $2$--designs \cite{Dankert2009}. One easy way to construct such a set is from the elements of a complete set of mutually unbiased basis (MUBs) \cite{Schwinger1960, Klappenecker2005}, which are only known to exist in Hilbert spaces whose dimension is the power of a prime number \cite{Ivonovic1981a,Wootters1989,Durt2010}. More recently, selective and efficient quantum process tomography (SEQPT) protocols which, by making use of complete sets of MUBs in auxiliary Hilbert spaces, can be easily implemented in any dimension, 
were presented in Ref.~\cite{Perito2018}, while its experimental utility has been shown on a photonic platform \cite{PearsStefano2021}.

All the protocols named above are designed to describe quantum processes that preserve the trace of the quantum input state but, in general, one has to deal with open quantum systems, where the evolution is not necessarily described by a trace-preserving quantum map. 
Even when the quantum process we intend to describe is,  theoretically, a trace-preserving map, real-world quantum channels or devices have inherent losses. Thus, a generalization of the previous schemes to reconstruct non trace-preserving processes is required. As an example, in Ref.~\cite{Zheng2017}, we can observe the need to resort to QPT methods for non trace-preserving maps, to characterize  quantum algorithms implemented in a four-qubit superconducting quantum processor. 
In other cases, the quantum algorithms are implemented in a probabilistic platform. This is the case of the Knill-Laflamme-Milburn linear-optic quantum computing scheme~\cite{Knill2001,Politi2009}, where the inherent probabilistic nature of the implementation makes the complete process, 
a non trace-preserving one \cite{Rahimi-Keshari2011,Shi2021}. 

In this work, we show that when some \textit{a priori} information about losses in a given channel is at hand, a generalization of the SEQPT protocols to reconstruct such a non trace-preserving channel, can be obtained. Additionally to a detailed description of the proposed scheme, that works for Hilbert spaces of arbitrary finite dimension, we test its validity by performing the experimental reconstruction of non trace-preserving quantum processes in dimensions $d=3$ and $d=6$, in the \texttt{ibmq\_lima} (a 5-qubit quantum processor) provided by IBM Quantum Services~\cite{ibmq}. To this end, we proposed an encoding that embeds the $d-$dimensional Hilbert space in the $n-$qubits based processor. Namely, we codified the quantum systems of dimension (qudits) for $d=3$ and $d=6$, into a two-qubits (dimension $2^2=4$) or a three-qubits system (dimension $2^3=9$), respectively, using the remaining subspace to introduce a controlled loss. While there are several examples of quantum algorithms  implemented in a qubit-based superconducting quantum processor  \cite{Zambrano2020, Satyajit2018, Behera2019}, including quantum tomography schemes and a recent SEQPT implementation for trace-preserving maps \cite{Gaikwad2022}, we show that these platforms are also suitable to implement and validate quantum algorithms and tasks for {\it qudits} of {\it arbitrary dimensions}.

The paper is organized as follows. In Section \ref{sec:formalism} we start by reviewing the standard formulation of the SEQPT (Subsection \ref{sec:standard_method}). In Subsection \ref{sec:seqpt-ntp}, we introduce the proposed generalization 
of the method when the process to be characterized does not preserve trace. The details of our experimental implementation in an IBM quantum computer, for different quantum processes acting on systems of dimensions $d=3$ and $d=6$, are presented in Section \ref{sec:ibmq-exp}. In Section \ref{sec:resultsanddiscussion} we show and discuss the obtained results, and finally we conclude with the outstanding aspects of the work in Section \ref{sec:conclusions}.

\section{Formalism}\label{sec:formalism}

We will start this section by briefly reviewing the formalism to describe a quantum process, followed by the standard formulation of the SEQPT protocol introduced in Refs.~\cite{Bendersky2008,Bendersky2009}. It is said that this is 
\textit{selective}, in the sense that it allows obtaining a particular coefficient of the process matrix $\chi$ without having to perform the full QPT, and \textit{efficient}, since such a coefficient can be determined with sub-exponential resources. Then, we will present its generalization to reconstruct non trace-preserving quantum processes.

A quantum process can be mathematically represented by a linear and completely positive map $\E$, from the set of density operators into itself~\cite{Nielsen}. The effect of this map on a quantum state $\rho$ can always be written as a Kraus decomposition $\mathcal{E} \left(\rho\right)=\sum_{k}A_k \rho
A_k^\dagger$, where $\{A_k\}_k$ is a set of linear operators that act on a Hilbert space $\mathcal{H}$, and
satisfy the relation $\sum_{k}A_k A_k^\dagger\leq\Id$ \footnote{The meaning of $\sum_{k}A_k A_k^\dagger\leq\Id$ is that all the eigenvalues of the $\sum_{k}A_k A_k^\dagger-\Id$ matrix are negative.}. This restriction implies that $0\leq \mathrm{Tr}\left[\mathcal{E}(\rho)\right]\leq 1$ for any $\rho$, which guaranties that $\mathcal {E}$ represents a \textit{physical} quantum process.
In order to relate the decomposition of the map
with measurable parameters, one can choose a convenient basis of operators $\left\lbrace E_i, i=0,\ldots,d^2-1\right\rbrace$, with $d$ the dimension of the quantum system, and write each operator $A_k$ in this basis. Therefore, the action of the process is expressed as
$\mathcal{E} \left(\rho\right)=\sum_{ij}\chi_j^iE_i
\rho E^j$,
where $\chi$ is a hermitian and positive $d^2\times d^2$ matrix,
and we have
adopted the convention $E^{j}\equiv E_{j}^\dagger$. Now, the trace condition
is given by the following inequality
\begin{eqnarray}\label{trace_condition}
\sum_{ij}\chi_j^iE^j E_i\leq \Id,
\end{eqnarray} 
and it is said that the evolution of the system, under the considered process, is described by a \textit{trace-preserving} or a \textit{non trace-preserving} map, depending on whether or not the equality is fulfilled. Hence, determining all the coefficients $\chi_j^i$ is equivalent to completely characterize the process.

\subsection{SEQPT for trace-preserving maps}\label{sec:standard_method}
Let us review the standard SEQPT protocol for {\it trace-preserving} maps. The key quantity for {\it selectively} reconstructing the process matrix $\chi$ is the {\it average survival probability}  $\bar{F}(\E_j^i)$, which can be defined as 
\begin{equation}\label{Fid_modified_channel}\bar{F}(\mathcal{E}_j^i)=\int_\mathcal H d\psi \; \expectation{\psi}{\E \left(E^i \ketbra{\psi}{\psi}E_j\right)}{\psi}\, ,
\end{equation}
and it is directly related to the coefficient $\chi_j^i$ by the following expression: 
\begin{equation}\label{eq:F_relate_Chi}
  \chi_j^i=\frac{d+1}{d} \bar{F}(\E_j^i)-\frac{1}{d^2}\, \Tr\left[(\sum_{\mu\nu}\chi_{\mu}^{\nu} E^{\nu}E_{\mu}) E^i E_j\right].
\end{equation}

\begin{figure}[t]
  \begin{center}
  \includegraphics[width=.4\textwidth]{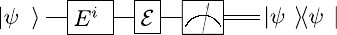}
  \caption{Circuit schematic for measuring the survival probability of the state $\ket{\psi}$ through the modified channel $\E_{i}^{i}$. By sampling over $\ket{\psi}$ in the state 2--design $X$, we can estimate the diagonal element $\chi_{i}^{i}$,  corresponding to the process matrix of $\E$.\label{fig:schematic-circuit-a}}
  \end{center}
  \end{figure}

Since the equality in Eq.~(\ref{trace_condition}) is fulfilled, the second term in the previous expression is reduced to $\frac{1}{d+1}\delta_j^i$ when the summation is performed, and we arrive to the uncoupled equation 
\begin{equation}\label{eq:chi}
  \chi_j^i=\bar{F}(\E_j^i)\frac{d+1}{d}-\frac{\delta_j^i}{d}\;,
\end{equation}
which allows any matrix coefficient $\chi_j^i$ to be related to the determination of a single average survival probability.

In order to experimentally estimate $\bar{F}(\E_j^i)$, the integral can be replaced with an average over a particular finite set of states known as  {\it uniform} 2--design \footnote{A state 2--design is a finite set of states
$X=\left\lbrace\ket{\psi_m}, m=1,...,N\right\rbrace$ over which the mean
value of any function $f$ that is quadratic in $P_\psi$ gives the
same mean value as on the set of all possible states in $\mathcal{H}$. That is, $\int_\mathcal H d\psi\,f\left(P_\psi\right)= \sum_{m=1}^N  p_m~f\left(P_{\psi_m}\right)$, where $\{p_m\}_{m=1,...,N}$ is some fixed probability distribution and the integration is performed over the Haar measure. In particular, if the probability distribution is uniform ($p_m = 1/\left| X \right|=1/N~\forall m=1,\dots, N$), $X$ is called \textit{uniform} 2--design.}, 
\begin{equation}\label{eq:F_2design}
  \bar{F}(\E_j^i)=\frac{1}{N}\sum_{m=1}^N  \expectation{\psi_m}{\E \left(E^i \Proj{\psi_m} E_j\right)}{\psi_m}.
\end{equation}
 Note that, the term in the right side of Eq.~(\ref{eq:F_2design}) has a clear experimental interpretation: it represents the survival probability through a modified channel $\E_j^i$, averaged over $X$. For example, the diagonal coefficient $\chi_i^i$ can be obtained by preparing all the states in $X$, one by one, and measuring their survival probability through the modified channel $\E_i^i$, as schematized in Fig.~\ref{fig:schematic-circuit-a}. In the non-diagonal case, the modified channel $\E_j^i$ is not physical, but the coefficient $\chi_j^i$ can still be obtained from the outputs of \textit{at most} four of that circuits (see Appendix \ref{ap:ancilla}).

 \begin{figure}[t]
	\begin{center}
	\includegraphics[width=.4\textwidth]{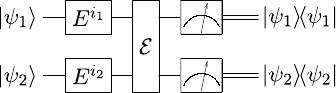}
	\caption{Circuit schematic to perform SEQPT when the dimension of the Hilbert space can be factorized as powers of two prime numbers. By sampling over the product 2--design $X_{\otimes}$, the survival probability of the state $\ket{\psi} = \ket{\psi_1}\otimes \ket{\psi_2}$ through the modified channel $\E_{i_1i_2}^{i_1i_2}$, we can estimate the diagonal element $\chi_{i_1i_2}^{i_1i_2}$,  corresponding to the process matrix of $\E$.\label{fig:schematic-circuit-b}}
	\end{center}
	\end{figure}

The problem of finding a state 2--design is easily solved when the dimension $d$ of the system is the power of a prime number. In this case, it is always possible to construct a complete set o mutually unbiased basis (MUBs) \cite{Wootters1989} which automatically constitutes an uniform 2--design \cite{Klappenecker2005}. In other cases is not trivial to compute the integral in Eq.~(\ref{Fid_modified_channel}) as an average over a finite set of states, and thus the protocol becomes impractical. However, the generalization of the MUBs-based SEQPT protocol to quantum processes in a Hilbert space of arbitrary dimension $d$, has been developed in Ref.~\cite{Perito2018}, and experimentally implemented in Ref.~\cite{PearsStefano2021}, for the case of trace-preserving maps. The approach exploits the fact that tensor products of 2-designs can be used to approximate a state 2-design\footnote{It should be noted that there are ways to obtain 2--designs which do not require the existence of a complete set of MUBs. However, in this work we exploit the well understood protocol based on MUBs.}. Then, since an arbitrary dimension $d$ can always be factorized into power of prime numbers, i.e., $d=p_1^{n_1}p_2^{n_2}\dots p_N^{n_N}$ with $\{p_i\}_{i=1}^N$ all different prime numbers, the tensor product of maximal sets of MUBs in Hilbert spaces of dimension $D_1=p_1^{n_1}$, $D_2=p_2^{n_2},\dots, D_N=p_N^{n_N}$, provides a good approximation for integration purposes.

To fix ideas, we will briefly describe the bipartite case, 
where the dimension of the Hilbert space is factorized as
$d=D_1 D_2$,  but the extension to any other multipartite case is straightforward~\cite{Perito2018}. A more thorough explanation, already presented in \cite{Perito2018}, can be followed in Appendix~\ref{ap:bipartite}. We start by expanding the map
$\mathcal E$ over $\mathcal{H}$ $(\mathcal{H} = \mathcal{H}_1 \otimes \mathcal{H}_2)$, in a basis that is a tensor product of operators
acting on $\mathcal{H}_1$ and $\mathcal{H}_2$, being $D_1=p_1^{n_1}=\dim(\mathcal{H}_1)$ and $D_2=p_1^{n_2}=\dim(\mathcal{H}_2)$,
respectively. This basis can be chosen as tensor products of two orthogonal operator  bases
$\{ E_{j_1j_2} \equiv E_{j_1} \otimes E_{j_2}
\}_{j_1=0,\dots,D_1^2-1}^{j_2=0,\dots,D_2^2-1}$, where each element
$E_{j_i}$ ($i=1,2$) is an unitary matrix.  Hence, we are able to rewrite $\mathcal E$ as
\begin{equation}
  \mathcal E (\rho) =  
\sum_{\mu_1\mu_2\nu_1\nu_2} \chi_{\nu_1\nu_2}^{\mu_1\mu_2} E_{\mu_1\mu_2} \rho   E^{\nu_1\nu_2}   
\, ,
\label{eq:canal2primos}
\end{equation}
for some coefficients $\chi_{\nu_1\nu_2}^{\mu_1\mu_2}$. Furthermore, we can easily define a uniform 2--design $X_1$ ($X_2$), based on MUBs, for the Hilbert space  $\mathcal{H}_1$ ($\mathcal{H}_2$). 

Now, the survival probability is evaluated, experimentally, for states belonging to $X_{\otimes} = \{\ket{\psi_1}\otimes\ket{\psi_2},\, \mathrm{for\ }\ket{\psi_1}\in X_1,\ \ket{\psi_2}\in X_2\}$, that is the set of all tensor products an element of $X_1$ and an element of $X_2$. Although the average with this sampling scheme does not directly yield $\fid(\E_{j_1j_2}^{i_1i_2})$, it allow us to estimate three related average quantities $\fid_{\otimes}(\E_{j_1j_2}^{i_1i_2})$, $\fid_1(\E_{j_1j_2}^{i_1i_2})$ and $\fid_2(\E_{j_1j_2}^{i_1i_2})$ (see Appendix \ref{ap:bipartite}), that can be exactly related to the average survival probability:

\begin{eqnarray}\label{eq:chifid_Bip_alt}
  \bar{F}(\E_{j_1j_2}^{i_1i_2})&=&\frac{1}{(d+1)}\{\fid_{\otimes}(\E_{j_1j_2}^{i_1i_2})(D_1+1)(D_2+1)\nonumber\\&+&\frac{2}{d}\Tr\left[(\sum_{\mu\nu}\chi_{\mu}^{\nu} E^{\nu}E_{\mu}) E^i E_j\right]\\
  &-&\fid_1(\mathcal E_{j_1j_2}^{i_1i_2})(D_1+1)-\fid_2(\E_{j_1j_2}^{i_1i_2})(D_2+1)\}.\nonumber
  \end{eqnarray}

An example circuit for this case is depicted in Fig.~\ref{fig:schematic-circuit-b}. This measurement correspond to the reconstruction of a given diagonal coefficient $\chi_{i_1i_2}^{i_1i_2}$. Here, an arbitrary state $\ket{\psi}=\ket{\psi_1}\otimes\ket{\psi_2}$ in $X_{\otimes}$ is prepared and its survival probability trough the modified channel $\E_{i_1i_2}^{i_1i_2}$ is obtained as output. Finally, as in the case of power of prime dimension, if a trace-preserving map is assumed, from Eq.~\eqref{eq:chifid_Bip_alt} an expression for any $\chi_{j_1j_2}^{i_1i_2}$, totally uncoupled to the others elements of the process matrix is obtained. 

\subsection{SEQPT formulation for non trace-preserving maps}\label{sec:seqpt-ntp}

The assumption of the trace-preserving property of the map $\mathcal{E}$ is essential to achieve \textit{selectivity} in the tomographic method. If we
drop this assumption, the Eq.~(\ref{eq:F_relate_Chi})
can no longer be decoupled to obtain Eq.~(\ref{eq:chi}). Thus, each matrix coefficient $\chi_{j}^i$ is coupled to
all the others, and of the order of $\sim d^4$ circuits like the one in Fig.~\ref{fig:schematic-circuit-a}, should be performed to solve the resulting equation system.
However, with some \textit{a priori} information about
the process to be characterized, this issue could be overcome. 
Indeed,  
it can be done by defining a semidefinite
positive Hermitian operator $\mathcal{P}$, given by
\begin{align}\label{eq:P_operator}
	\mathcal{P} = \sum_{ij}
	\chi_j^i E^j  E_i.
\end{align}
This operator was originally introduced in Ref.~\cite{Bongioanni2010}, in the context of standard quantum process tomography (SQPT) for non trace-preserving maps.
It encodes the losses of the system, so that, given an input state $\rho$, the probability of obtaining an output state after the process $\E$ (probability of
success of the process), is
\begin{equation}\label{eq:p-trace}
    \Tr\left[\E(\rho)\right] = \Tr\left[\mathcal{P}\rho\right].
\end{equation}

The meaning of $\mathcal{P}$ becomes clear when it is analyzed in its diagonal form: let us
write $\mathcal{P} = \sum_i \gamma_i\ketbra{\gamma_i}{\gamma_i}$, where $\ket{\gamma_i}$
are the eigenstates and $0 \leq \gamma_i\leq 1$ the corresponding
eigenvalues. Hence, the state $\ket{\gamma_i}$ has a probability of success equal to $\gamma_i$. 
For a trace-preserving map $\gamma_i=1\; \forall i$, and
$\mathcal{P}=\Id$. Otherwise, we have a non trace-preserving process with state-independent ($\gamma_i=\gamma<1,\; \forall i$) or state-dependent (at least one $\gamma_i<1$ and different to the others)
success probability. 
For example, in a photonic experiment, this decrease in the trace value may be associated
to a global loss, when $\mathcal{P}=\gamma\Id$, or to different losses in each path of the setup, in another case. Then, provided that we can previously have a description of $\mathcal{P}$, which is possible in many realistic scenarios, Eq.~(\ref{eq:F_relate_Chi}) is now reduced to
\begin{equation}\label{eq:tracesNTP}
  \bar{F}(\E_j^i)=\frac{d^2\chi_j^i+\Tr[\mathcal{P}E^i E_j]}{d(d+1)} \, ,
\end{equation}
and any element $\chi_j^i$ can be independently computed from the corresponding average fidelity $\bar{F}(\E_j^i)$. In the particular case in which $\mathcal{P}$ is a multiple of the identity operator, Eq.~(\ref{eq:tracesNTP}) returns to Eq.~(\ref{eq:chi}) but modified by an additive factor $\frac{(1-\gamma)}{d}\delta_j^i$. So that, this derivation also includes trace-preserving maps where $\gamma=1$.

In the bipartite case, the uncoupled expression for the element $\chi_{j_1j_2}^{i_1i_2}$ is (the complete deduction is presented in Appendix \ref{ap:ntp-bipartite}):
\begin{eqnarray}
	\label{eq:chifid-p}	\chi_{j_1j_2}^{i_1i_2}&=&\fid_\otimes(\E_{j_1j_2}^{i_1i_2})\frac{(1+D_1)(1+D_2)}{d}\\\nonumber
	&+& \frac{1}{d^2} \Tr\left(\mathcal{P} E^{i_1i_2} E_{j_1j_2} \right)\\\nonumber
	&-&\fid_1(\mathcal E_{j_1j_2}^{i_1i_2})\frac{(1+D_1)}{d}-\fid_2(\E_{j_1j_2}^{i_1i_2})\frac{(1+D_2)}{d}.\nonumber
\end{eqnarray}

In general, a possible shortcoming of a non-trace preserving formulation is that a previous knowledge of the $\mathcal{P}-$matrix is needed. However, there are several use cases where that requirement is available. For example, when the losses of each individual building block of a quantum circuit are characterized, $\mathcal{P}$ could be estimated {\it a priori} for each particular configuration of the full circuit. Therefore, the implementation of the SEQPT scheme is suitable to certify the correct implementation and the performance of the algorithm of interest (see for example \cite{Zheng2017}). In a similar way, the implementation of non-deterministic quantum gates \cite{Knill2001,Lund2002,Politi2009} could be tested by the method that we present in this work. 

It should be mention that, if the reconstruction of the quantum map were carried out assuming, incorrectly, that the trace is preserved, the fidelity would drop with respect to the reconstruction assuming that the trace is not preserved (see our results in Section \ref{sec:resultsanddiscussion} B), and this difference tends to increase with losses \cite{Bongioanni2010}. 
Furthermore, the topic of QPT schemes for non trace-preserving maps is seldom treated in the literature. The the standard quantum process tomography (SQPT) method works off the shelf \cite{Bongioanni2010,Bouchard2019} but lacks efficiency, given that it requires a number of measurements that grows exponentially with the dimension of the system. Other approaches are: i) direct characterization of quantum dynamics \cite{Wang2007}, which requires an ancillary state, ii) the use of coherent states \cite{Rahimi-Keshari2011}, and iii) QPT via weak values \cite{Kim2018}. On that regard, the generalization of SEQPT to non trace-preserving processes enables the possibility to perform efficient and selective tomography for this type of maps.

Finally, it is important to mention that if we have no knowledge of $\mathcal{P}$, Eq. \ref{eq:p-trace} can be used to gain partial information about it, by measuring the probability of success in some selected basis. For example, by just sampling states of the $2-$design, all the diagonal elements of the $\mathcal{P}$-matrix can be estimated. We think that, as future work, it is worth exploring whether this partial information of $\mathcal{P}$ can be combined with optimization methods, or even with variations of SEQPT 
that simultaneous estimate any diagonal coefficient of the process matrix $\chi$ and not a single one (Generalization I in Ref. \cite{Bendersky2009}), to perform a selective and efficient tomography even in absence of information about $\mathcal{P}$.

\section{Experimental realization in an IBM 
    Quantum Computer}\label{sec:ibmq-exp}

To experimentally validate the  
non trace-preserving formulation 
of the SEQPT method, we used the 
superconducting quantum computer \texttt{ibmq\_lima}.
This is a freely available quantum processor of $5$ qubits, that can be
programmed with the open source Python framework, Qiskit
\cite{qiskit}. We implemented the SEQPT to characterize a non trace-preserving quantum process for both $d=3$, a dimension with a known $2-$design, and $d=6$, the minimal dimension in which the bipartite extension is non trivial.

\subsection{Reconstruction of quantum processes in $d=3$}

To represent a $3-$dimensional quantum system (qutrit) in a qubit based quantum computer, we have used $2$ qubits of the
processor, with the following assignation:\begin{eqnarray}\label{eq:2qubit-3dstate}
  \ket{00} \equiv \ket{0}_3 &~&\;\;\;\;\;\; \ket{10} \equiv \ket{2}_3 \nonumber\\
  \ket{01} \equiv \ket{1}_3 &~& \;\;\;\;\;\underbrace{\boxed{\ket{11} \equiv \ket{\ell}}}_{\mathrm{lost\ state}},
\end{eqnarray}
where $\mathcal{B}=\{\ket{q_1q_0}\}_{q_j=0,1}$ is the computational basis for the $2-$qubit system, and $\mathcal{B}'=\{\ket{k}_3\}_{k=0,\dots,2}$ is the canonical basis for the $3-$dimensional Hilbert space $\mathcal{H}^{(3)}$. It is important to mention that, since there is one more state in the basis of a $2-$qubit system, this extra state should be discarded. We take advantage of this 
to simulate losses in the qutrit evolution. That is, any unitary transformation that couples the subset $\{\ket{00},\ket{01},\ket{10}\}$ to the remaining element $\ket{11}\in \mathcal{B}$, will represent a non trace-preserving process on the Hilbert space defined by $\mathcal{B}'$.

The target qutrit process, $\E^{(3)}$, to be implemented and reconstructed, is based in a Hadamard gate extended as a qutrit quantum operation~\cite{Di2011}:
\begin{equation}H_{01} = \frac{1}{\sqrt{2}} 
  \begin{pmatrix} 
    1 &  1 & 0 \\
    1 & -1 & 0 \\
    0 &  0 & \sqrt{2}
  \end{pmatrix}.\label{eq:h01-matrix}
\end{equation}
As can be seen, $H_{01}$ maps the basis states $\ket{0}_3\rightarrow \frac{\ket{0}_3+\ket{1}_3}{\sqrt{2}}$ and $\ket{1}_3\rightarrow \frac{\ket{0}_3-\ket{1}_3}{\sqrt{2}}$, and leaves state $\ket{2}_3$ unchanged. However, since we are interested in non trace-preserving process, we have included a \emph{beamsplitter-like loss} of $50\%$ affecting the state $\ket{2}_3$, which results in a Kraus decomposition with a single unitary operator, $\E^{(3)}(\rho)=A^{(3)}\rho\; (A^{(3)})^\dagger$, where
\begin{IEEEeqnarray*}{rCclc}\nonumber A^{(3)} \;=   &\frac{1}{\sqrt{2}}& \;(\ketbra{0}{0} - \ketbra{1}{1}+ \ketbra{2}{2}\\\IEEEyesnumber
  & & +\;\ketbra{0}{1} + \ketbra{1}{0} ). 
\end{IEEEeqnarray*}
\begin{figure}[h]
  \includegraphics[width=1\linewidth]{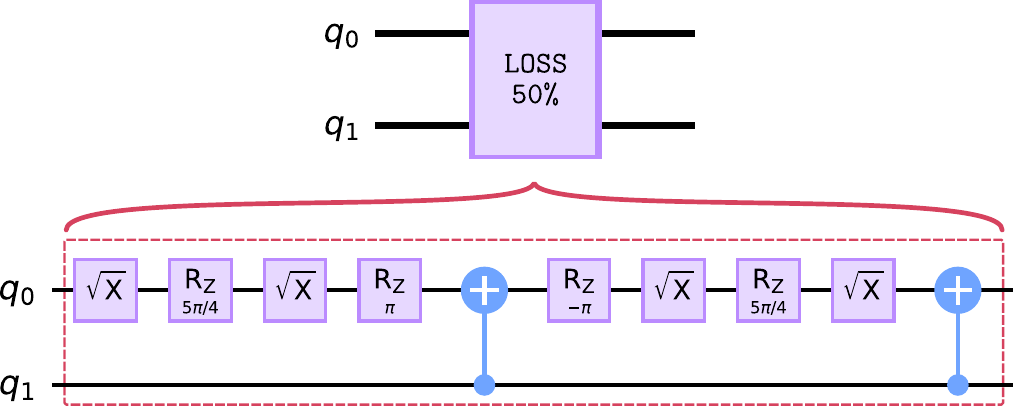}
  \caption{Circuit to implement a beamsplitter-like loss affecting the state $\ket{10}\equiv\ket{2}_3$. This state is coupled to the discarded 2-qubit state $\ket{11}\equiv\ket{\ell}$, with a $50\%$ probability. \label{fig:beamsplitter-circuit}}
\end{figure}
Hence, losses in the qutrit evolution will correspond to a coupling between the state $\ket{2}_3$ and the discarded 2-qubit state renamed as $\ket{\ell}$. In the basis $\mathcal{B}$, this coupling can be performed by a controlled-Hamadard gate 
on the target qubit 0, controlled by qubit 1. \figref{fig:beamsplitter-circuit}, shows the equivalent circuit that implement such a loss in the IBMQ processor (\texttt{LOSS 50\%}). The percentage of loss, $r\times100$, can be set to a different value by changing the angle of rotation of some of the $\mathrm{R}_z$--gates:
this allows simulating non trace-preserving channels, with arbitrary losses, without increasing the depth of the circuit. The schematic of the circuit to implement a \textit{general} beamsplitter-like gate is discussed in Appendix \ref{ap:beamsplitter-circuit}.
\begin{figure}[t]
  \includegraphics[width=.95\linewidth]{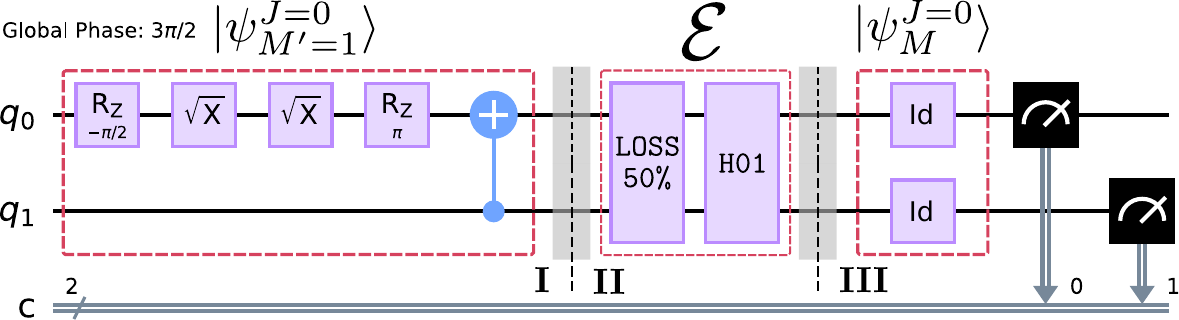}
  \caption{Example of a circuit that samples an element of the $2-$design $X$, in a Hilbert space of dimension $d=3$, trough the modified channel $\E_i^i$. ($\mathbf{I}$) In this particular case, the sampled element, after being modified by the corresponding $E_i$ operator, is $\ket{\psi_{M'=1}^{J=0}}$. ($\mathbf{II}$) The process $\E$ corresponds to a \texttt{LOSS} $\texttt{50\%}$ followed by the operation \texttt{H01}. ($\mathbf{III}$) In the measurement stage, an unitary transformation (the identity in this example) represents the change of basis from $\mathcal{B}_{J}$ to the canonical one ($\mathcal{B}_{J=0}$), followed by the measurement of the state of each qubit.  \label{fig:example-circuit-d3}}
\end{figure}
Then, it is easy to see that in the basis $\mathcal{B}'$, the loss operator $\mathcal{P}$ corresponding to the process $\E^{(3)}$ has a diagonal matrix form:
\begin{eqnarray}
\mathcal{P} =
  \begin{pmatrix} 
    1 &  0 & 0 \\
    0 & 1 & 0 \\
    0 &  0 & \frac{1}{2}
  \end{pmatrix}, \label{eq:p-matrix-d3}
\end{eqnarray}
that is, $\gamma_0=\gamma_1=1$ and $\gamma_2=r=\frac{1}{2}$. This is the information that we will consider (\emph{a priori}) known to reconstruct the process. 

At this point, we will describe our implementation of the SEQPT protocol in the 2--qubit processor. As example, Fig.~\ref{fig:example-circuit-d3} shows the schematic circuit, which closely resemble the one in Fig.~\ref{fig:schematic-circuit-a}, to sample a particular element of the 2--design $X$ through the modified channel $\E^i_i$. As a state 2--design we choose a complete set of MUBs, that in $d=3$ has exactly 12 elements.
The first stage of the circuit in Fig.~\ref{fig:example-circuit-d3} corresponds to the preparation of the selected input state $\ket{\psi_m}\equiv \ket{\psi_M^J}$, modified by the operator $E^i$, where $J$ $(J=0,\dots,d)$ indicates the MUB and $M$ $(M=0,\dots,d-1)$ refers to a particular state in that $J$--MUB. 
Due to the selected basis of operators, the resulting state is an element of the same $J$--MUB: $\ket{\psi_{M'}^J}$ (see Appendix~\ref{ap:bases}). The second stage corresponds to the circuit that implement the 
process $\mathcal{E}^{(3)}$, while the last stage performs a measurement in the $J$--MUB, required to estimate the survival probability of the sampled state (Eq.\eqref{eq:F_2design}). In the presented example, as the projection is on the canonical basis $\mathcal{B}_{J=0}$, there is no need for a change of basis prior to detection of each qubit. 

\subsection{Reconstruction of quantum processes in $d=6$}

To represent a $6-$dimensional quantum system (qudit),
at least 3 qubits of the IBMQ computer are needed. 
Furthermore, to implement the bipartite version of the SEQPT we must factorize $\mathcal{H}^{(6)}$ as $\mathcal{H}_1 \otimes \mathcal{H}_2$, with $d = D_1 D_2 = 2 \times 3$.
A suitable alternative is to consider a similar assignation to the one in the Eq. \eqref{eq:2qubit-3dstate} to expand $\mathcal{H}_2$, and add an extra qubit for $\mathcal{H}_1$:
\begin{eqnarray}\label{eq:3qubit-6dstate}
  &&\;\ket{000}\equiv\ket{0}_6 = \ket{0}_2\otimes\ket{0}_3   \;\;\;\;\;\;  \ket{100}\equiv \ket{3}_6 = \ket{1}_2\otimes\ket{0}_3 \nonumber\\
  &&\;\ket{001}\equiv \ket{1}_6 = \ket{0}_2\otimes\ket{1}_3 \;\;\;\;\;\;  \ket{101}\equiv \ket{4}_6 = \ket{1}_2\otimes\ket{1}_3 \nonumber\\
  &&\;\ket{010}\equiv \ket{2}_6=\ket{0}_2\otimes\ket{2}_3  \;\;\;\;\;\;  \ket{110}\equiv \ket{5}_6=\ket{1}_2\otimes\ket{2}_3 \nonumber\\
  &&\;\;\;\;\;\;\;\;\;\;\;\;\;\;\;\;\;\;\;\;\underbrace{\boxed{\ket{011}\equiv\ket{\ell_1}  \;\;\;\;\;\;\;  \ket{111}\equiv\ket{\ell_2} }}_{\mathrm{lost\ states}}\;,
\end{eqnarray}
where $\mathcal{B}''=\{\ket{q_2q_1q_0}\}_{q_j=0,1}$ is the computational basis for the $3-$qubit system, and $\mathcal{B}'''=\{\ket{k}_6\}_{k=0,\dots,5}$ the canonical basis for $\mathcal{H}^{(6)}$. It is worth noting that, in this case, we have used two states ($\ket{011}$ and $\ket{111}\in\mathcal{B}''$ renamed as $\ket{\ell_1}$ and $\ket{\ell_2}$), to simulate losses in the evolution of the qudit. For this purpose, the same circuit of Fig. \ref{fig:beamsplitter-circuit} was programmed to implement a beamsplitter-like loss with a $50\%$ chance of coupling $\ket{2}_6$ and $\ket{5}_6$ to the lost states $\ket{\ell_1}$ and $\ket{\ell_2}$, respectively.

The target process that we have implemented for its subsequent reconstruction, can be decomposed as $\E^{(6)}(\rho)=A^{(6)}\rho\;(A^{(6)})^\dagger$, with
\begin{IEEEeqnarray*}{rCclc}\nonumber A^{(6)}&  = & &\phantom{(} \ketbra{0}{0} + \ketbra{3}{3} \phantom{)} & \\\IEEEyesnumber
  & + & \; e^{i\pi/3} \;& \left(\ketbra{1}{1} + \ketbra{4}{4}\right) & \\
  & + & \frac{1}{\sqrt{2}}& \left( \ketbra{2}{5} +  \ketbra{5}{2}\right).&
\end{IEEEeqnarray*}
This corresponds to a phase shift of $\varphi=\pi/3$ affecting the states $\ket{1}_6$ and $\ket{4}_6$, a swap operation between the states $\ket{2}_6$ and $\ket{5}_6$, and losses that also affect this two states, \texttt{SWAP25 + LOSS 50$\%$}. 
For this non trace-preserving process, the diagonal form of the $\mathcal{P}$-matrix is 
\begin{eqnarray}
  \mathrm{diag} (\mathcal{P}) =  
  \begin{pmatrix} 
    1 & 1 & \frac{1}{2} & 1 & 1 & \frac{1}{2}
  \end{pmatrix}, \label{eq:p-matrix-d6}
\end{eqnarray}
that is, $\gamma_i=1$ for $i\in \{0,1,3,4\}$ while $\gamma_i=r=\frac{1}{2}$ for $i\in \{2,5\}$.
\begin{figure*}[ht!]
  \includegraphics[width=0.75\textwidth]{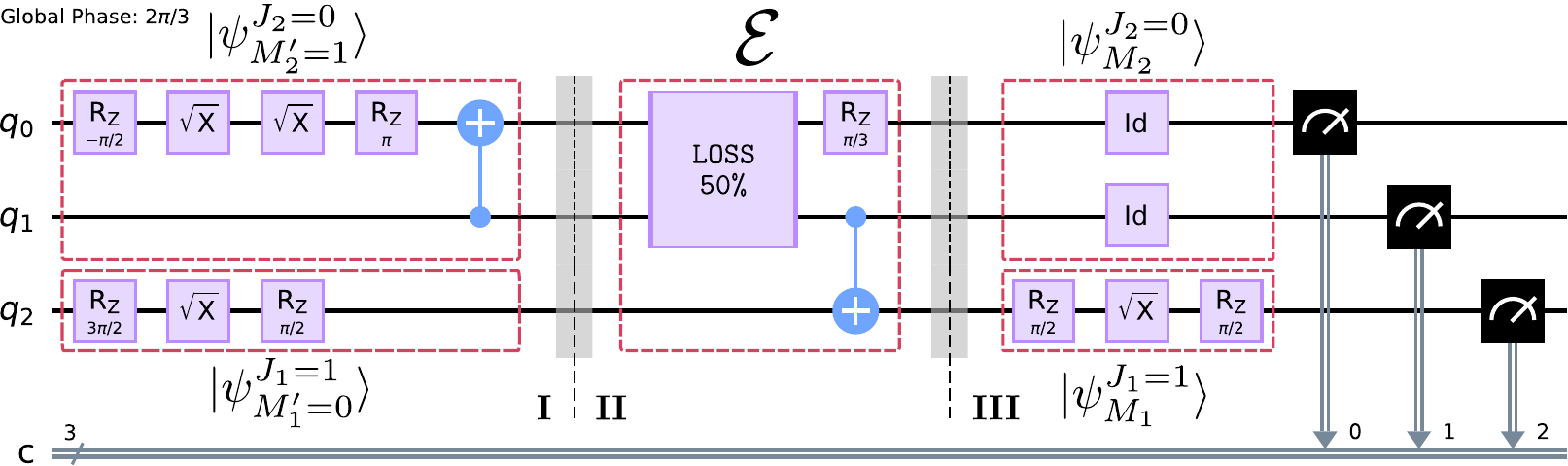}
  \caption{Example of a circuit that samples an element of the product of $2-$designs $X_\otimes=X_1\otimes X_2$
    in a Hilbert space of dimension $d=6$, through the modified channel $\E^{i_1i_2}_{i_1i_2}$. The 
    subspace $\mathcal{H}_1$, which dimension is $D_1=2$, coincides with the third qubit.  ($\mathbf{I}$) The sampled element, after being affected by the operator $E_{i_1i_2}$, is $\ket{\psi_{M'=0}^{J_1=1}} \otimes \ket{\psi_{M'=1}^{J_2=0}}$. ($\mathbf{II}$) The process $\E$ corresponds to a \texttt{LOSS} $\texttt{50\%}$ followed by the $\mathrm{R}_z$ and the \texttt{CNOT} gates that implement a \texttt{SWAP25}. ($\mathbf{III}$) Finally, an unitary transformation is applied to change of basis from $\mathcal{B}_{J_1=1} \otimes \mathcal{B}_{J_2=0}$ to the canonical one ($\mathcal{B}_{J_1=0} \otimes \mathcal{B}_{J_2=0}$), followed by the measurement of each qubit.  \label{fig:example-circuit}}
\end{figure*}

Finally, with the encoding election for a qudit  (Eq.~\eqref{eq:3qubit-6dstate}), the SEQPT protocol in the 3–qubit processor is implemented by programming the schematic circuit shown in Fig.~\ref{fig:example-circuit}, which resembles the circuit for the bipartite case depicted in Fig. \ref{fig:schematic-circuit-b}. In this example, an element $\ket{\psi_m}\equiv\ket{\psi_{M_1}^{J_1}} \otimes \ket{\psi_{M_2}^{J_2}}$ in the product $2-$design $X{_\otimes}$, is sampled trough the modified channel $\E^{i_1i_2}_{i_1i_2}$ and projected to estimate its survival probability (Eq.~\eqref{mean_fidelity} in Appendix \ref{ap:bipartite}). Both states 2--design, $X_1$ and $X_2$, were chosen to be a complete sets of MUBs, so that $\ket{\psi_{M_1}^{J_1}}$ and $\ket{\psi_{M_2}^{J_2}}$ are the $M_1$ and $M_2$ elements of the $J_1$--MUB in $D_1=2$, and the $J_2$--MUB in $D_2=3$, respectively, giving a total of 72 elements in $X_\otimes$.
In the first stage, the state $\ket{\psi_m}$ is prepared and modified by the operator $E^{i_1i_2}$ which, analogously to the case $d=3$ gives, by construction, another element of $X_{\otimes}$. Thus, the experimental implementation of the modified channel only requires the preparation of elements in $X_{\otimes}$ as input states of the channel $\mathcal{E}^{(6)}$, which corresponds to the second stage of Fig.~\ref{fig:example-circuit}, while the last stage performs a projective measurement on the tensor product of ${J_1}$ and ${J_2}$ bases.

It is important to mention that, as we will see in Section~\ref{sec:resultsanddiscussion}, the particular choice of MUBs as state 2--designs, together with the operator bases used in this work, results in a significant reduction of the number of circuits to be implemented to carry out the SEQPT. An explicit construction of the 2--designs and the operator bases are discussed in Appendix~\ref{ap:bases}. 

 \begin{figure}[b]
        \includegraphics[width=.85\linewidth]{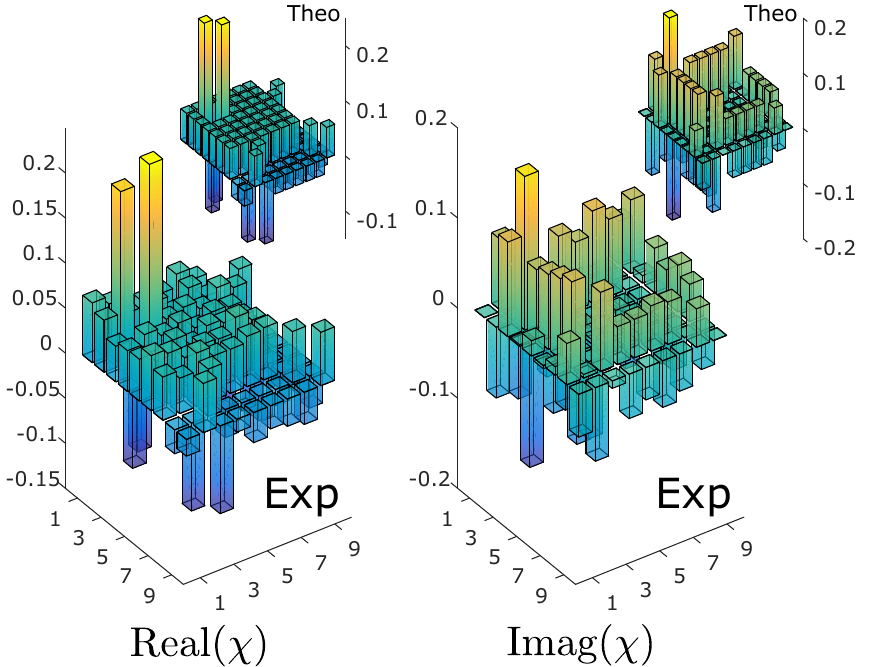}
        \caption{Comparison of the real (left panel) and imaginary (right panel) parts of the matrix $\chi_{\mathrm{exp}}$, experimentally obtained trough SEQPT and a convex optimization. This is the matrix that characterizes the process $\E^{(3)}$. Upper insets show the corresponding values of the target matrix $\chi_{\mathrm{theo}}$.\label{fig:rho-ibm-d3-alt}}
        \end{figure}

\section{Results and Discussion} \label{sec:resultsanddiscussion}

In this section we will show the main experimental results that were obtained when using the SEQPT method
in the reconstruction of non trace-preserving quantum maps. They correspond to quantum circuits with losses for two different dimensions: $d=3$ (power of prime dimension) and $d=6$ (non trivial example of arbitrary dimension). In the case of the $d=6$ we also explore, in detail, the selective and efficient properties of the method.

\subsection{Process matrix reconstruction: case $d=3$}\label{subsec:results:qutrit}
To assess the viability of the SEQPT in the case of non trace-preserving quantum maps, we first used the method to perform a full tomography of the target channel $\E^{(3)}$. To this end, all of the $9\times9=81$
coefficients of the expected theoretical process matrix $\chi_\mathrm{theo}$ were reconstructed, by sampling for each coefficient, the 12 elements in the state 2--design $X$. Due to the particular relation between the selected operator basis and the $2-$design (Appendix \ref{ap:bases}), many of the circuits needed to perform that sampling are repeated. Thus, the full tomographic reconstruction requires the implementation of only 36 different circuits like the one in Fig.\ref{fig:example-circuit-d3}, out of a total of $81\times12\times4=3888$ circuits. This shows the importance of an adequate choice of the basis of operators for a given state 2--design, in order to avoid redundancy.
Each circuit was repeated 8192 times (\emph{shots}) to estimate the frequency distribution of the outcomes, and thus infer the corresponding survival probabilities $\bar{F}(\E_j^i)$. Since an important source of error in the circuit evaluation comes from the measurement stage, an error mitigation routine was used to post-process the frequency distribution. This routine, \texttt{ignis.mitigation.measurement}, is provided by the Qiskit framework module, and relies in an initial calibration measurement, performed on the superconducting quantum computer, of the qubit state prepared in the computational basis. 

As in most tomographic methods, due to the effect of statistical and systematic errors,
the reconstructed channels are not physical \cite{james2005measurement,Knee2018}. For that reason, after a first estimation of the matrix $\chi_{\mathrm{theo}}$ we solved the following convex optimization problem\footnote{The optimization problem was solved trough the Python library \texttt{cvxpy} \cite{cvxpy}.}:
\begin{equation}
\begin{aligned}
  \min_{\chi_\mathrm{opt}}\quad & \left\lVert \chi_{\mathrm{raw}} - \chi_\mathrm{opt} \right\rVert_F\\
  \mathrm{s.t.} \quad & \chi_\mathrm{opt}\geq 0,\\
  &0\leq\Tr\left(\sum_{ij} {\chi^i_j}_\mathrm{opt} E^j E_i\right) = \Tr(\mathcal{P})\leq d,
  \end{aligned}\label{eq:optim}
\end{equation}
where $\chi_\mathrm{raw}$ is the process matrix experimentally reconstructed using the SEQPT method, and $\chi_\mathrm{opt}$ the resulting optimized matrix. This optimization is similar to the one used in Ref. \cite{Gaikwad2022}, where the optimized matrix is the closest matrix, in the sense of the Frobenious norm, to the one estimated by the tomographic method, subject to the constraints of physicality. These constraints are: i) the matrix must be \emph{semi definite positive}, which in turn implies that the quantum map is complete positive \cite{Knee2018}; ii) the condition 
on the trace, that ensures a non-increasing trace map, whose value must be equal to the trace of the $\mathcal{P}$-matrix, \emph{a priori} known. For lossless evolution we have $\mathcal{P}\equiv \Id$, and the last constraint reduces to the usual one for a trace-preserving quantum map, $\Tr\left(\sum_{ij} {\chi_j^i}_\mathrm{opt} E^j E_i\right) = d$. 

\figref{fig:rho-ibm-d3-alt} shows the bar plots representing the real and imaginary part of the reconstructed process matrix, $\chi_\mathrm{exp}\equiv\chi_{\mathrm{opt}}$ for the channel \texttt{HO1+LOSS} $\texttt{50\%}$. For comparison reasons, we also show the real and imaginary part of $\chi_\mathrm{theo}$ (see upper insets in Fig.~\ref{fig:rho-ibm-d3-alt}). As a figure of merit of the reconstruction process we used the normalized process fidelity  
\begin{equation}\label{eq:fidelity}
  \mathcal{F}\equiv \mathcal{F}(\chi_\mathrm{theo},\chi_\mathrm{exp})=\frac{\Tr\left[\sqrt{\sqrt{\chi_\mathrm{theo}}\chi_\mathrm{exp}\sqrt{\chi_\mathrm{theo}}}\right]}{ \sqrt{\Tr\left[\chi_\mathrm{theo}\right]} \sqrt{\Tr\left[\chi_\mathrm{exp}\right]}}.
\end{equation} 
This definition acts as a geometric distance in the space of density matrices and, due to the Choi-Jamiolkowsi isomorphism \cite{Mohseni2008}, it can be used as a geometric distance measure between quantum processes. It should also be noted, that the normalization factor is included so that it is $\mathcal{F}=1$ for identical processes, even when they are non trace-preserving \cite{Bongioanni2010, Zheng2017}. 

The resulting fidelity value for the process $\E^{(3)}$ is $\mathcal{F}_\mathrm{SEQPT} = 0.956$. For completeness, we have also reconstructed the process under SQPT method. 
Briefly, the SQPT that was conducted on the IBMQ superconducting quantum computing consists of preparing $\sim d^2$ states which are a combination of states in the $d$--dimensional canonical basis: $\ket{k}$, $\ket{k} + \ket{k'}$ and $\ket{k} + i \ket{k'}$, with $k=0,\dots, d-1$ and $k'=(k+1),\dots, d-1$. Each of this states is then reconstructed, after being affected by the quantum process, trough a standard quantum state tomography (QST). The circuits were implemented by using the \texttt{state\_tomography\_circuits} function, provided by the Qiskit Ignis library. As the QST method provided by this framework works for $n$-qubit states, we then projected the resulting reconstructed state onto the $3$--dimensional subspace of interest. Finally, the linear relation between the prepared and reconstructed states is inverted to obtain the process matrix. The experimental estimation of $\chi_{\mathrm{theo}}$ under the SQPT resulted in a process fidelity $F_\mathrm{SQPT} = 0.974$, which is slightly higher than that obtained with the SEQPT. However, it should be noted that the implementation of SQPT required 81 circuits, in contrast with the 36 circuits required by our SEQPT implementation.

To better qualify the performance of the proposed method (SEQPT) for the reconstruction of non trace-preserving maps, 
\begin{table}[h!]
  \centering
  \begin{tabular}{c|l|c|c|c|c}
    \hline 
    \multicolumn{2}{c|}{\multirow{2}{*}{\;Process fidelity\;\;}} &  \multicolumn{2}{c|}{TP process}& \multicolumn{2}{|c}{NTP process}\\ \cline{3-6}
   \multicolumn{2}{c|}{} & \;SEQPT\;& \;SQPT\; & \;SEQPT\; & \;SQPT\;\;\; \\ \hline \hline
   \multirow{3}{*}{$\;\;\;d=3\;\;\;$} &\;\texttt{ID} & 0.971 & 0.982 & 0.958 & 0.964\\ 
                           &\;\texttt{H01} & 0.970 & 0.981 & 0.956 &  0.974\\ 
                           &\;\texttt{H12} & 0.953 & 0.973 & 0.934 &  0.954\\ \hline  
   \end{tabular}
 \caption{Summary of the fidelities of reconstruction for different non trace-preserving  processes (NTP) and their trace-preserving counterpart (TP) in $d=3$.}\label{tb:results}
\end{table}
\begin{figure*}[ht!]
    \includegraphics[width=.85\linewidth]{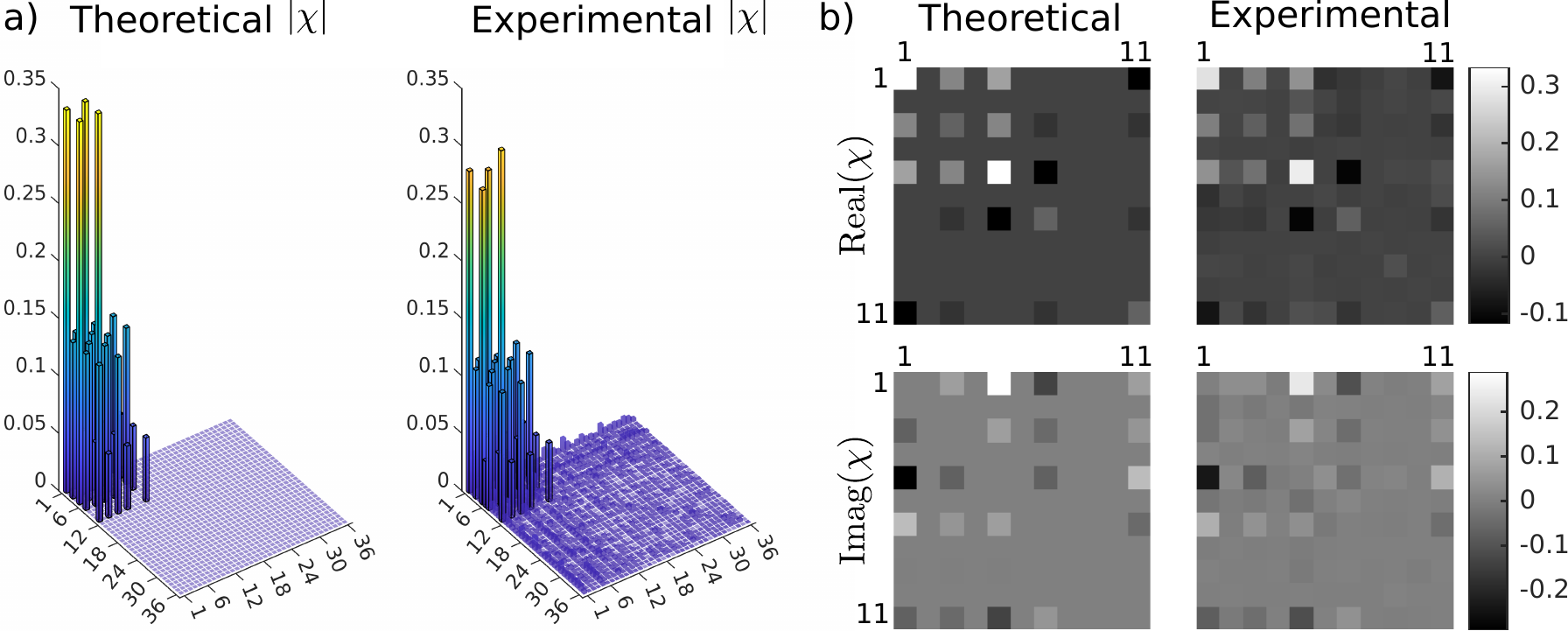}
    \caption{\textbf{a)} Comparison of the absolute values of the elements
      of $\chi_\mathrm{theo}$ and $\chi_\mathrm{exp}$ for the NTP
      process $\mathcal{E}^{(6)}$, measured in a 5-qubit computer from the IBMQ
      network. \textbf{b)} Detail of the the first 11 basis elements. This
      elements correspond to the non-zero block in the theoretical matrix
      (upper-left block in the theoretical plot of a)). The grey-level map
      shows the real and imaginary part for the theoretical and the
      experimental reconstructed matrices.\label{fig:rho-ibm-d-six}}
    \end{figure*}
in comparison with the standard accepted one (SQPT), and also to understand the origin of possible errors that are later reflected in fidelity values below 1, we have reconstructed the trace-preserving version of $\E^{(3)}$, where the 50\% loss subprocess was removed. In addition,
other simple processes for $d=3$, both trace-preserving and non trace-preserving, were reconstructed: \texttt{ID} (the identity process), \texttt{ID+LOSS} (a 50\% loss coupled to the identity process), the qutrit Hadamard gate \texttt{H12}, and its version with losses \texttt{H12+LOSS} (see Appendix \ref{ap:krauss-processes} for the explicit mathematical description of these processes). 
Table \ref{tb:results} compares the values of the fidelity of reconstruction for the studied $3$--dimensional quantum processes. As to highlight, the reconstruction of non trace-preserving processes through the SEQPT differs by less than $3\%$ from reconstruction using SQPT. Besides, the reconstruction of non trace-preserving processes results in a slightly lower fidelity than the reconstruction of their trace-preserving counterpart (in absence of the \texttt{LOSS} subprocess). This drop in the fidelity value could be attributed to the additional number of gates needed to implement the loss of the target process.

\subsection{Process matrix reconstruction: bipartite case $d=6$}\label{subsec:result:bipartite}
The advantages of the SEQPT method with respect to the SQPT becomes more evident in higher dimensions. In that regard, we will experimentally study the reconstruction of the channel $\E^{(6)}$. In this dimension ($d=6$), we can also assess the viability of the non trace-preserving version of the SEQPT for the general case, when the dimension of the Hilbert space is not a power of a prime number.

We start by performing the full reconstruction of the selected target channel, in the IBMQ processor. For this, the $36\times36=1296$ coefficients of the $\chi_{\mathrm{theo}}$ matrix were estimated by sampling all the elements in the tensor product of 2--design $X_\otimes=X_1\otimes X_2$. As in the case $d=3$, due to the particular relation between the selected basis of operator and the product of $2-$design, many of the circuits to perform the sampling are repeated. Thus, with this choice, the full reconstruction under the SEQPT method requires the implementation of only 432 different circuits, like the one in Fig.~\ref{fig:example-circuit}, out of a total of $\sim36\times36\times72\times4=373248$ that certainly would lead in redundancy of measurements. Finally, the statistics over 8192 repetitions of each circuit is collected to obtain the required survival probabilities $\bar{F}_{\otimes}(\E^{i_1i_2}_{j_1j_2})$.
\begin{figure}[h!] \includegraphics[width=.95\linewidth]{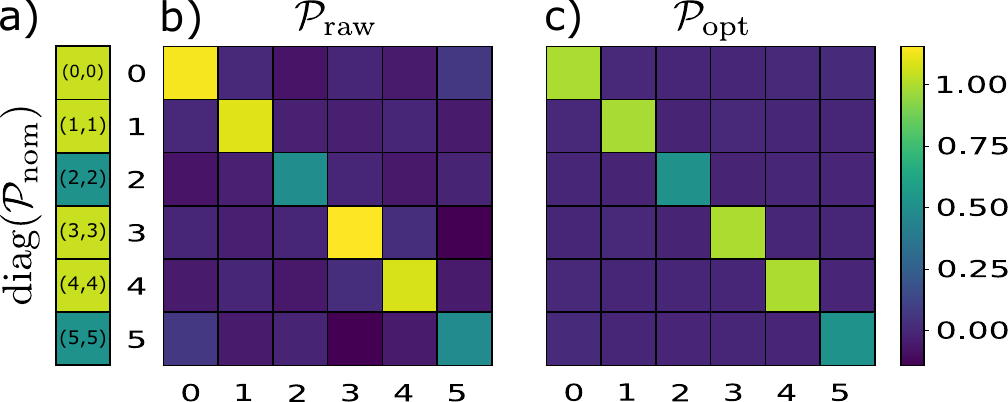}
  \caption{Comparison between the diagonal form of the nominal $\mathcal{P}$-matrix and the experimentally reconstructed one. \textbf{a)}  Diagonal of the $\mathcal{P}$-matrix for the nominal process $\E^{(6)}$. \textbf{b)} Full $\mathcal{P}$-matrix obtained for the process reconstructed by the NTP version of the SEQPT method.  \textbf{c)} $\mathcal{P}$-matrix for the process post-processed by the convex optimization routine. 
  \label{fig:p-matrices}}
\end{figure}

The bar plots in Fig. \ref{fig:rho-ibm-d-six} a) indicate the absolute values of the coefficients of the theoretical matrix $\chi_\mathrm{theo}$ (left) and the experimentally reconstructed one $\chi_\mathrm{exp}$ (right). Besides, in Fig. \ref{fig:rho-ibm-d-six} b) we show a grey-level map detail of the $11\times11$ nonzero elements in $\chi_\mathrm{theo}$, comparing the real (upper panels) and imaginary (lower panels) parts of its coefficients (left panels) with the corresponding coefficients of $\chi_\mathrm{exp}$ (right panels). In this case, we obtained a value for the process fidelity $\mathcal{F}_\mathrm{SEQPT} = 0.913$, while for the same process reconstructed with SQPT, which requires a total of 972 circuits, we obtained a slightly lower fidelity, $\mathcal{F}_\mathrm{SQPT} = 0.871$.  Besides, 
if the process is assumed to be trace-preserving, the standard SEQPT method results in a reconstruction fidelity of only $0.838$.

Given that the \emph{a priori} information of the $\mathcal{P}$--matrix was used to experimentally estimate, individually, any coefficient of the matrix $\chi_{\mathrm{theo}}$ of the non trace-preserving target process $\E^{(6)}$, it is important to check that the obtained results are compatible with that initial assumption. In Fig.~\ref{fig:p-matrices} a), we show a color map representing the values of the coefficient of the $\mathcal{P}$--matrix in its diagonal form, and in Fig.~\ref{fig:p-matrices} b), the corresponding values obtained after the full reconstruction from Eq.~\eqref{eq:P_operator}. It can be seen that some elements in the diagonal are slightly greater than 1 (up to $14\%$), which indicates that the process is not physical. Even so, there is a clear similarity with the $\mathcal{P}$--matrix of the target process. Finally, in Fig.~\ref{fig:p-matrices} c), we show the $\mathcal{P}$--matrix obtained from the post-processed data, i.e., after convex optimization (see Eq.~\eqref{eq:optim}). Although the constraint in the optimization stage is on $\Tr(\mathcal{P})$, the value of each of its elements is very close to that of the nominal process (less than $4\%$ of difference).
\begin{table}[h!]
  \centering
  \begin{tabular}{c|l|c|c|c|c}
    \hline 
    \multicolumn{2}{c|}{\multirow{2}{*}{Process fidelity}} &  \multicolumn{2}{c|}{TP process}& \multicolumn{2}{|c}{NTP process}\\ \cline{3-6}
   \multicolumn{2}{c|}{} & \;SEQPT\;& \;SQPT\; & \;SEQPT\; & \;SQPT\;\;\; \\ \hline \hline
      \multirow{3}{*}{$\;\;\;d=6\;\;\;$} & \;\texttt{ID} & 0.950 & 0.947 & 0.930 & 0.938\\ & \;\texttt{PHASE} & 0.949 & 0.945 & 0.930&0.939\\ & \;\texttt{SWAP25} & 0.906 & 0.908 & 0.913&0.871\\ \hline 
   \end{tabular}
 \caption{Summary of the fidelities of reconstruction for different non trace-preserving  processes (NTP) and their trace-preserving counterpart (TP) in $d=6$.}\label{tb:results6}
\end{table}

For a wider analysis of performance, we also implemented and experimentally reconstructed the trace-preserving counterpart of $\E^{(6)}$, where the loss subprocess was removed, and other simple process in $d=6$:
\texttt{PHASE} (a phase shift of $\pi/3$ on the elements $\ket{1}$ and $\ket{4}$) and \texttt{PHASE+LOSS} (similar to \texttt{PHASE}, but with an additional 50\% loss). The explicit Kraus operators for these processes are presented in Appendix \ref{ap:krauss-processes}. Comparing both reconstruction methods, the difference in the fidelity values is, in all these cases, less than $1\%$.

\subsection{Efficiency}
As was mention in Section~\ref{sec:intro}, a remarkable feature of the SEQPT method is that it is both \emph{selective} and \emph{efficient}. This makes the protocol ideally suited for reconstructing quantum channels that have only a few non-zero elements in its matrix $\chi$. For example, if $\chi$ has $K$ non-zero elements, it is possible to reconstruct only these elements (\textit{selectivity}), and with a number of evaluation circuits that scales with $K$, independently of the dimension $d$ of the Hilbert space (\textit{efficiency}) \cite{Perito2018}. In contrast, the SQPT method reconstructs the entire matrix for which it requires a number of circuits that quickly grows with the dimension as $\sim d^4$.
\begin{figure}[ht]
\includegraphics[width=.95\linewidth]{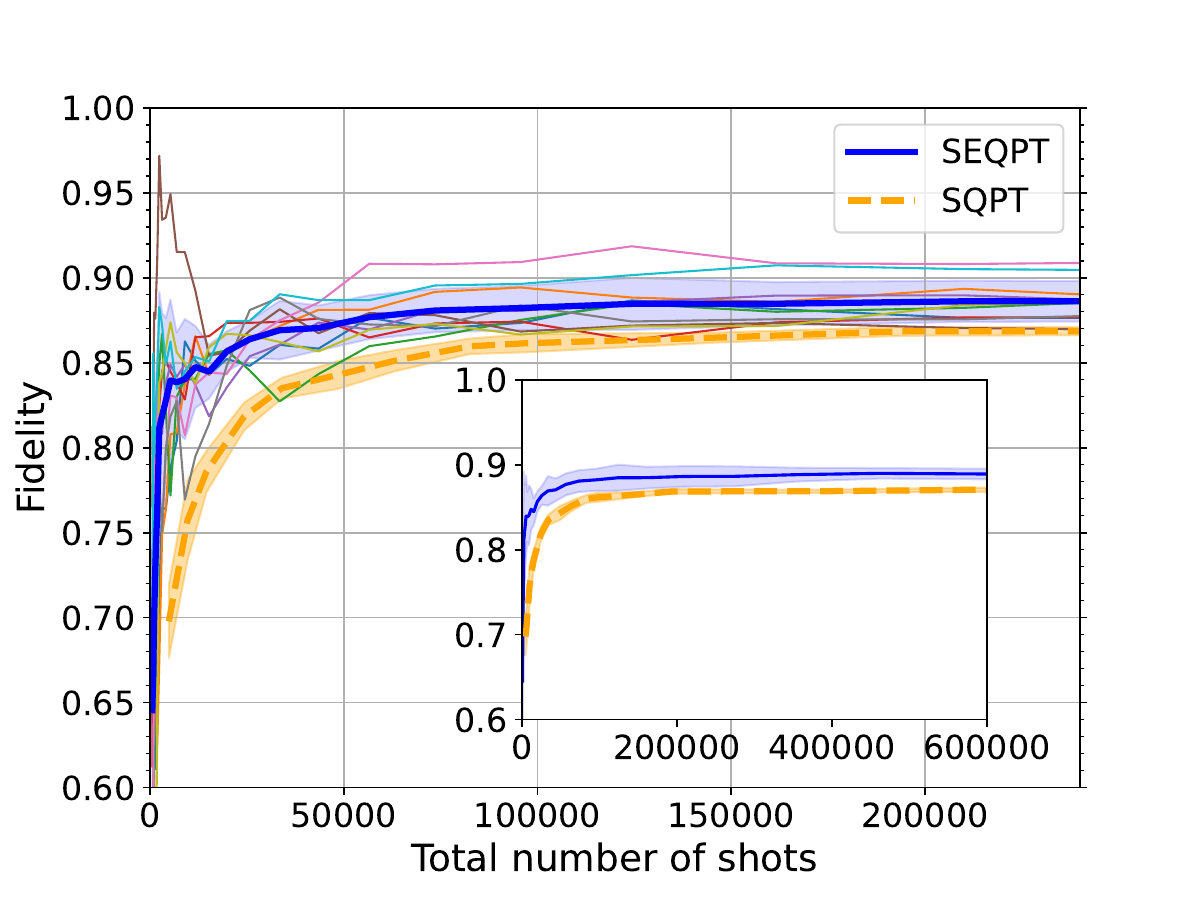}
  \caption{Process fidelity between the nominal quantum process $\mathcal{E}^{(6)}$ and the reconstructed one, for an increasing number of shots. Each thin solid line
    corresponds to a different random sampling of elements in $X_\otimes$. The sampling is performed to reconstruct the 25 non-zero elements of the target process matrix $\chi_{\mathrm{theo}}$, from SEQPT. The thick solid line
    corresponds to the mean process fidelity among those 10 different samplings, while the shaded area represents one standard
    deviation. For comparison, mean process fidelity and standard deviation for 10 realizations of the SQPT full reconstruction are displayed.\label{fig:efficiency}}
\end{figure}

To perform an efficient reconstruction of the matrix element $\chi^i_{j}$ one must appeal to the fact that the mean fidelities $\bar{F}_\otimes(\E^i_{j})$ and $\bar{F}_k(\E^i_{j})$ ($k=1,2$) can be estimated by
averaging the 
survival probabilities of a random subset of states in $X_\otimes$,
through the modified channel $\E^i_{j}$, instead of performing the complete sum in Eqs. \eqref{mean_fidelity}-\eqref{eq.Fid1}. At each sampling step $m$, an element $\ket{\psi_{m}}$ of $X_\otimes$ is randomly chosen (with replacement), the corresponding circuit to estimate the survival probabilities of this quantum state is executed once (one \emph{shot}), and the outcome is used to update the values of $\bar{F}_\otimes(\E^i_{j})$ and $\bar{F}_k(\E^i_{j})$. These estimates improve iteratively as the number of sampling steps $m$ grows.
In this way, we have carried out the reconstruction of the 25 non-zero elements of the target matrix $\chi_{\mathrm{theo}}$, which we previously reconstructed completely  (Fig.~\ref{fig:rho-ibm-d-six}).

Figure \ref{fig:efficiency} shows the process fidelity $\mathcal{F}$ between the target and the reconstructed process, as a function of the total number of shots for both the SEQPT method and the traditional method SQPT. Each thin line represents a realization of this SEQPT protocol, i.e., the reconstruction of the 25 non-zero elements of $\chi_{\mathrm{theo}}$ from a particular random sampling of $m$ elements in $X_\otimes$ ($m\leq \num{240e3}$, or$\;\sim$\num{10e3} \emph{shots} per non-zero matrix element). 
The thick solid line and shaded area represent the mean value of $\mathcal{F}_\mathrm{SEQPT}$ resulting from 10 realizations, and its standard deviation, respectively. To compute this fidelity, the remaining elements of the experimental matrix, not measured in this case, were assumed to be zero in accordance with what was expected theoretically, and the resulting process matrix, $\chi_{\mathrm{raw}}$, was optimized to ensure its physicality according to Eq.~\eqref{eq:optim}. The inset in Fig. \ref{fig:efficiency}, displays the results from a larger sampling ($m$ up to \num{600e3}, or \num{24e3} samples for each non-zero matrix element). 

On the other hand, the thick dashed line and shaded area in Fig.
\ref{fig:efficiency} represent the mean value of $\mathcal{F}_\mathrm{SQPT}$ for the SQPT method, resulting from 10 realizations, and its standard deviation, respectively. For a given realization, each of the 972 required circuits is executed an increasing number of times, and the corresponding fidelity $\mathcal{F}_\mathrm{SQPT}$ is updated with the increasing statistic.

In the SEQPT reconstruction we can see that the mean fidelity value increases quickly with the number of sampled states, reaching a stable value of $0.889$. Ideally, this value should be $\sim 1$, if only statistical errors are considered. However, the readout fidelity of the \texttt{ibmq\_lima} quantum computer,
which accounts for the readout error, was estimated to be only $0.938$. It should be also noted that  
the limit value that reaches the mean fidelity is slightly lower than the fidelity value obtained when a full reconstruction of the process was done by the same method ($\mathcal{F}_{\mathrm{SEQPT}}=0.913$). As it was described in Section \ref{subsec:results:qutrit}, in that case the readout error could be mitigated, after a calibration in the canonical basis, by post-processing the distribution probability of the outcomes of each circuit. In the present case, we ran each circuit only one time: since an individual shot does not represent the complete probability distribution for the circuit, thus it cannot be mitigated with the standard routine provided by the Qiskit framework module. In the future, an iterative scheme, where the probability distribution of each circuit is estimated with every new shot, could be envisioned to improve the reconstruction quality.

Finally, we can see that with an $m$ up to $\num{30e3}$ ($\sim$ \num{1200} samples per non-zero matrix element) is enough for a process fidelity whose value differs by $2.5 \%$ of its stable value. For comparison, in the SQPT at least \num{50e3} total shots were needed to achieve a fidelity value within $2.5 \%$ of its stable value. To contextualize this difference, for such a precision the total running time in the IBMQ processor is $\SI{9}{\sec.}$ for the SEQPT, i. e., it is a \SI{67}{\percent} faster than SQPT, which takes $\SI{15}{\sec}$. And what is more important,
in sharp contrast to the SQPT, the execution time of the SEQPT method will not increase when the dimension $d$ increases, since to a fix precision, $m$ should be increased only if the number of non-zero elements in the process matrix is greater \cite{Schmiegelow2011,Perito2018}.

\section{Conclusions}\label{sec:conclusions}

In this work, we have presented a generalization of the SEQPT protocol for non trace-preserving maps. The proposed method, that works for arbitrary dimensions, uses the \emph{a priori} information of the loss matrix to reconstruct individual elements of the process matrix.  To test this scheme, we realized an experimental implementation of the method in a five-qubit superconducting IBM quantum processor. We have successfully reconstructed several processes both in prime-power dimension ($d=3$) where a state 2--desing is known, and in dimensions where this does not occur ($d=6$). 

On the one hand, we have shown that it is possible to efficiently reconstruct non trace-preserving processes, with high precision, within the readout error of the current quantum computers. Since quantum processors are very sensitive to the environment, we have to deal with noisy devices, which makes it relevant to have methods for the reconstruction of quantum processes, that save resources and have the ability to account for non trace-preserving stages.

On the other hand, the implementation of such processes in a superconducting quantum processor was made possible by using the discarded Hilbert subspace to introduce controlled losses,  
generalizing the type of quantum processes that can be implemented and test in these computers.
In addition, this shows a way to use qubit-based quantum processor as test benches for quantum circuits in dimensions that are different of $2^N$ ($N\in \mathcal{N}$), that is, for \textit{qudit} spaces of arbitrary dimension $d$.

\begin{acknowledgments}
We acknowledge Prof. Claudio Iemmi for helpful discussions  about the focus of this work. This work was supported by Universidad de Buenos Aires
(UBACyT Grant No.~20020170100564BA). Q.P.S. was supported
by a CONICET Fellowship. We also acknowledge the use of IBM Quantum services for this work. The views expressed are those of the authors, and do not reflect the official policy or position of IBM or the IBM Quantum team. 
\end{acknowledgments}

\appendix

\section{Modified channel and survival probabilities}\label{ap:ancilla}

\begin{figure}[H]
  \begin{center}
  \includegraphics[width=.4\textwidth]{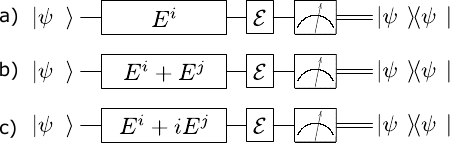}
  \caption{Schematic of the circuit for measuring the survival probability of the state $\ket{\psi}$ through the modified channel $\E_{i}^{i}$ (panel a)) or $\E_{j}^{i}$ (panels b) and c)). By sampling these circuits over $\ket{\psi}$ in the state 2--design $X$, we can estimate any element $\chi_{j}^{i}$, corresponding to the process matrix of $\E$.\label{figApp:schematic-circuit}}
  \end{center}
  \end{figure}

In order to make the protocol experimentally clear, we will describe how to interpret the action of the modified channel $\E_{j}^{i}$ in terms of the  physical channel $\E$. It should be noted that, regardless of the peculiarities of the experimental setup, this is a general description of the circuits that must be implemented to obtain the coefficients of the process matrix $\chi$.
\\\

i. \emph{Diagonal case:}\ for $i=j$, the effect of the modified channel $\E_{i}^{i}$
on a given state
$\ket{\psi}$, is
\begin{eqnarray}
\E_{i}^{i}(\ketbra{\psi}{\psi}) &=& \E(E^{i} \ketbra{\psi}{\psi}E_{i})
=\E(E^{i}P_{\psi}E_{i}),
\end{eqnarray}
and the probability that $\ket{\psi}$ survives $\E_{i}^{i}$ can be obtained by a projective measurement onto $\ket{\psi}$. This procedure is implemented by the circuit described in Fig.~\ref{figApp:schematic-circuit} a), whose output estimate the value of $\Tr[P_{\psi}\E(E^{i} P_{\psi}E_{i})]$.\\

ii. \emph{Non-diagonal case:\ }for $i\neq j$, the resulting modified channel is non-physical. In fact, its effect on a given state $\ket{\psi}$ corresponds to
\begin{eqnarray}
\E_j^i(\ketbra{\psi}{\psi}) = \E(E^i \ketbra{\psi}{\psi}E_j)
=\mathcal{E}(\ketbra{\alpha}{\beta}),
\end{eqnarray}
with $\ket{\alpha} = E^i\ket{\psi}$ and
$\ket{\beta} = E_j\ket{\psi}$. This is equivalent to the action
of the original channel $\mathcal{E}$ on the matrix
$\ketbra{\alpha}{\beta}$, which is not a density matrix, and therefore
does not represent a physical state. However, this matrix can always
be expressed as a linear combination of \textit{at most} four matrices, each corresponding to one projector. If $\ket{\alpha}$ and $\ket{\beta}$ are
orthonormal,
$\mathcal{E}(\ketbra{\alpha}{\beta}) = \mathcal{E}(\Proj{+}) +
\mathcal{E}(\Proj{-}) - \frac{1+i}{2} \left(\mathcal{E}(\Proj{\alpha})
  + \mathcal{E}(\Proj{\beta})\right),$ with
$\ket{+} = (\ket{\alpha} + \ket{\beta})/\sqrt{2}$ and
$\ket{-} = (\ket{\alpha} + i \ket{\beta})/\sqrt{2}.$ Even if they were not
orthonormal, a similar decomposition exists.  Then, the linearity of
$\E$ ensures that we can compute the action of the modified channel $\E_j^i$
over any state $\ket{\psi}$ as a linear combination of the action of the original channel $\mathcal{E}$
over a suitable choice of pure states. This implies that an additional operation must be performed on the input state $\ket{\psi}$, as sketched in Fig.~\ref{figApp:schematic-circuit} b) and Fig.~\ref{figApp:schematic-circuit} c).

Finally, the extension to the bipartite case, where the action of the modified channel on the state
$\ket{\psi}=\ket{\psi_1}\otimes\ket{\psi_2}\in X_{\otimes}$, is 
\begin{eqnarray}
\E_{j_1j_2}^{i_1i_2}(\ketbra{\psi}{\psi}) &=& \E(E^{i_1i_2} \ketbra{\psi}{\psi}E_{j_1j_2})\nonumber\\
&=&\mathcal{E}(\ketbra{\alpha}{\beta}),
\end{eqnarray}
being $\ket{\alpha} = E^{i_1i_2}\ket{\psi}=E^{i_1}\ket{\psi_1}\otimes E^{i_2}\ket{\psi_2}$ and
$\ket{\beta} =E^{j_1j_2}\ket{\psi}= E^{j_1}\ket{\psi_1}\otimes E^{j_2}\ket{\psi_2}$,
follows from i) if $i_1=j_1$ and $i_2=j_2$, or ii) if $i_1\neq j_1$ or $i_2\neq j_2$, but considering circuits of the type shown in Fig.~\ref{fig:schematic-circuit-b}.

\section{Average survival fidelities estimation in the bipartite case}\label{ap:bipartite}

In the bipartite case, instead of estimating directly the average survival probability $\bar{F}(\E_{j_1j_2}^{i_1i_2})$, we first estimate
\begin{eqnarray}\label{mean_fidelity}
  \bar F_\otimes(\E_{j_1j_2}^{i_1i_2})=
  \frac{1}{\left | X_{\otimes}\right |}
  \sum_{\ket{\psi}\in X_{\otimes}} \expectation{\psi}{\E_{j_1j_2}^{i_1i_2} \left(\Proj{\psi}\right)}{\psi}
  \end{eqnarray}
The magnitude $\bar F_\otimes(\E_{j_1j_2}^{i_1i_2})$ is the average survival probability over the tensor product of 2--design $X_\otimes\equiv\{\ket{\psi_1}\otimes\ket{\psi_2}/\ \mathrm{for}\ \ket{\psi_1}\in X_1,\, \ \ket{\psi_2}\in X_2\ \}$, and $|X_\otimes|$ represents the number of elements in the set. 

As shown in Ref.~\cite{Perito2018}, the integral that defines to $\bar{F}(\E_{j_1j_2}^{i_1i_2})$ can be expressed as:
\begin{eqnarray}\label{eq:chifid_Bip}
\bar{F}(\E_{j_1j_2}^{i_1i_2})&=&\frac{1}{(d+1)}\{\fid_{\otimes}(\E_{j_1j_2}^{i_1i_2})(D_1+1)(D_2+1)\nonumber\\&+&\frac{2}{d}\Tr\left[(\sum_{\mu\nu}\chi_{\mu}^{\nu} E^{\nu}E_{\mu}) E^i E_j\right]\\
&-&\fid_1(\mathcal E_{j_1j_2}^{i_1i_2})(D_1+1)-\fid_2(\E_{j_1j_2}^{i_1i_2})(D_2+1)\},\nonumber
\end{eqnarray}
where the value of $\fid_{1}(\E_{j_1j_2}^{i_1i_2})$ and $\fid_{2}(\E_{j_1j_2}^{i_1i_2})$, can be seen as the reduced survival probability, averaged over the state 2--design $X_{1}$ and $X_{2}$, respectively:  
\begin{eqnarray}\label{eq.Fid1}
&&\fid_k(\E_{j_1j_2}^{i_1i_2})=\\
&&\frac{1}{ D_{k'}\left| X_k \right |}
\sum_{\ket{\psi_k}\in X_k}\Tr\left[\left(P_{\psi_k}\otimes \Id_{k'}\right)\E_{j_1j_2}^{i_1i_2}\left(P_{\psi_k}\otimes \Id_{k'}\right)\right],\nonumber
\end{eqnarray}
where $k,k'=1,2$ and $k\neq k'$.

Thus, the {\it selectivity} of the method is given by the fact that a particular element $\chi_{j_1j_2}^{i_1i_2}$ can be determined by calculating the three mean fidelities $\fid_\otimes$, $\fid_1$ and $\fid_2$, over the modified channel $\E_{j_1j_2}^{i_1i_2}$. 
\section{Derivation of the SEQPT protocol for non trace-preserving maps in the bipartite case $X_{\otimes}=X_1\otimes X_2$}\label{ap:ntp-bipartite}

Let us consider a product Hilbert space $\hilb \equiv \hilb_1 \otimes \hilb_2$ of dimension $d$. We start by defining the average over the Haar measure of the product of two operators $A$ and $B$ as:
\begin{equation}
\langle A , B \rangle \equiv \int_{\hilb} d\psi \Tr \left[ P_{\psi} A P_{\psi} B \right] \, ,
\label{eq:inthaar}
\end{equation}
where $P_{\psi} \equiv \ket{\psi}\bra{\psi}$ and the integral is performed using the Haar measure over $\hilb$. We also define the product average
\begin{equation}
\langle A , B \rangle_{\otimes} \equiv \int_{\hilb_1}\int_{\hilb_2} d\psi_1 d\psi_2 \Tr \left[ P_{\psi_1\psi_2} A P_{\psi_1\psi_2} B \right] \, ,
\label{eq:inthaar_tensor2}
\end{equation}
where $P_{\psi_1\psi_2} \equiv P_{\psi_1} \otimes P_{\psi_2} = \ket{\psi_1}\bra{\psi_1} \otimes \ket{\psi_2}\bra{\psi_2}$,
and the integrals are performed over the Haar measure of each subsystem. Besides, the reduced averages are defined as:
\begin{equation}
\langle A , B \rangle_1 \equiv \int_{\hilb_1} d\psi_1 \Tr \left\{ P_{\psi_1} \Tr_2 \left[ A \left( P_{\psi_1}\otimes\frac{\mathbb{I}_2}{D_2} \right) B \right] \right\} \, ,
\label{eq:inthaar_1}
\end{equation}
\begin{equation}
\langle A , B \rangle_2 \equiv \int_{\hilb_2} d\psi_2 \Tr \left\{ P_{\psi_2} \Tr_1 \left[ A \left( \frac{\mathbb{I}_1}{D_1}\otimes P_{\psi_2} \right) B \right] \right\} \, ,
\label{eq:inthaar_2}
\end{equation}
where, for $k=1,2$, $\mathbb{I}_k$ indicates the identity operator acting on $\hilb_k$, $\Tr_k$ is the partial trace over subsystem $\hilb_k$, $D_k$ is the dimension of $\hilb_k$  and the integrals are performed using the Haar measure of the corresponding Hilbert space.

On the one hand, we will take into account the expression developed in Refs. \cite{Perito2018,PeritoPHd}, which relates the four averages defined above: 
\begin{equation}
\begin{split}
\langle A,B \rangle =&\frac{1}{d+1}  \left[  \left(D_1+1\right)\left(D_2+1\right)\langle A,B \rangle_\otimes  + \frac{2}{d}\Tr \left[ A B \right]\right. \\  
& \!\!\!\!\!\!\!\!\!\!\!\!\!\!\left.\phantom{\frac{2}{D}}- (D_1+1)\langle A , B \rangle_1 - (D_2+1) \langle A , B \rangle_2 \right] \, .
\end{split} 
\label{eq:funcionalproductovsoriginal}
\end{equation}

On the other hand, we will consider the average fidelity of a channel $\mathcal{E}$, defined as
\begin{equation}
\bar{F} (\mathcal{E}) \equiv \int_\hilb d\psi \Tr \left[P_\psi\;\mathcal{E}(P_\psi)  \right] \, ,
\label{eq:fidcanal}
\end{equation}
as well as the average product fidelity
\begin{equation}
\bar{F}_\otimes (\mathcal{E}) \equiv \int_{\hilb_1} \int_{\hilb_2} d\psi_1 d\psi_2 \Tr \left[ P_{\psi_1\psi_2} \mathcal{E} (P_{\psi_1\psi_2}  ) \right] \, ,
\label{eq:prodfidcanal}
\end{equation}
and the average reduced fidelities
\begin{equation}
\bar{F}_1 (\mathcal{E}) \equiv \int_{\hilb_1} d\psi_1 \Tr \left\{ P_{\psi_1} \Tr_2 \left[ \mathcal{E} \left( P_{\psi_1} \otimes \frac{\mathbb{I}_2}{D_2} \right)  \right] \right\}  \, , 
\label{eq:reducedfidcanal_1}
\end{equation}
\begin{equation}
\bar{F}_2 (\mathcal{E}) \equiv \int_{\hilb_2} d\psi_2 \Tr \left\{ P_{\psi_2} \Tr_1 \left[ \mathcal{E} \left( \frac{\mathbb{I}_1}{D_1}  \otimes P_{\psi_2} \right)  \right] \right\} \, .
\label{eq:reducedfidcanal_2}
\end{equation}
Recalling the expansion in Eq. \eqref{eq:canal2primos}, and given that each average $\langle \, , \, \rangle_\xi$ is bilinear in its arguments, we can write
\begin{equation}
\bar{F}_\xi (\bigeps_{j_1 j_2}^{i_1 i_2})  =\sum_{\substack{\mu_1\mu_2 \\ \nu_1\nu_2}} \chi_{\nu_1\nu_2}^{\mu_1\mu_2} \langle E^{i_1i_2} E_{\mu_1 \mu_2}  , E^{\nu_1\nu_2} E_{j_1j_2} \rangle_\xi \, ,
\end{equation}
where $\xi$ codifies which of the four fidelities defined in Eqs.~\eqref{eq:fidcanal}-\eqref{eq:reducedfidcanal_2}, and which of the four averages in Eqs.~\eqref{eq:inthaar}-\eqref{eq:inthaar_2}, we are referring to. From this identity and the relation in Eq.~\eqref{eq:funcionalproductovsoriginal}, it is obtained:
\begin{equation}
\begin{split}
\bar{F} (\bigeps_{j_1 j_2}^{i_1 i_2}) = \frac{1}{d+1} & \left[ (D_1+1)(D_2+1) \bar{F}_\otimes (\bigeps_{j_1 j_2}^{i_1 i_2})  \right.  \\ & +\frac{2}{d}\Tr\left(\mathcal{P} E^{i_1i_2} E_{j_1j_2} \right)  \\ &- (D_1+1) \bar{F}_1 (\bigeps_{j_1 j_2}^{i_1 i_2}) \\
&\left. - (D_2+1) \bar{F}_2 (\bigeps_{j_1 j_2}^{i_1 i_2}) \right] \, .
\end{split}
\label{eq:fidelidadesbipartito}
\end{equation}
Resorting to Eq.~\eqref{eq:tracesNTP}, the left side on the last equation can be related to the matrix coefficients to arrive to the desired result:
\begin{eqnarray}\label{eqApp:elements}
	\chi_{j_1j_2}^{i_1i_2}&=&\fid_\otimes(\E_{j_1j_2}^{i_1i_2})\frac{(1+D_1)(1+D_2)}{d}\\\nonumber
	&+& \frac{1}{d^2} \Tr\left(\mathcal{P} E^{i_1i_2} E_{j_1j_2} \right)\\\nonumber
	&-&\fid_1(\mathcal E_{j_1j_2}^{i_1i_2})\frac{(1+D_1)}{d}-\fid_2(\E_{j_1j_2}^{i_1i_2})\frac{(1+D_2)}{d}.\nonumber
\end{eqnarray}

Finally, bearing in mind that each fidelity on the right side of Eq.~\eqref{eqApp:elements} is quadratic in $P_{\psi_1}$, $P_{\psi_2}$ or $P_{\psi_1\psi_2}$, and defined as a Haar integral over Hilbert spaces with dimensions $D_1$ and $D_2$, when such dimensions are powers of a prime number, they can be computed by averaging the integrand over the corresponding $2$--designs $X_1$ or $X_2$, or over the tensor product 2--design $X_{\otimes}=X_1\otimes X_2$, respectively (see Eqs.~\eqref{mean_fidelity} and \eqref{eq.Fid1}).

\section{Beamsplitter-like circuit for simulating losses}\label{ap:beamsplitter-circuit}
\begin{figure}[h]
  \includegraphics[width=1\linewidth]{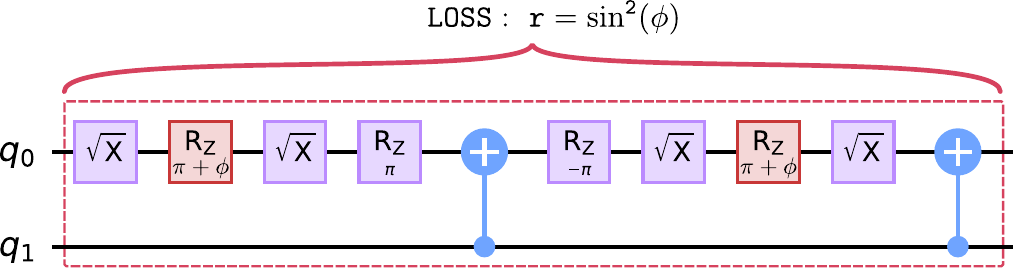}
  \caption{Circuit to implement a general beamsplitter-like loss affecting the state $\ket{10}\equiv\ket{2}_3$. A phase $\pi + \phi$ in both of the red-colored $\mathrm{R}_z$--gates results in a loss  $r=\sin^2(\phi)$ (beamsplitter reflectivity).  \label{fig:general-beamsplitter-circuit}}
\end{figure}

The loss implemented by the beamsplitter-like gate, which is shown in Fig.~\ref{fig:beamsplitter-circuit}, can be parameterized through the angle of rotation of the $\mathrm{R}_z$--gates. Figure \ref{fig:general-beamsplitter-circuit} shows the most general form of such a circuit. An argument of $\pi + \phi$ in both of the red colored $\mathrm{R}_z$--gates results in the equivalent unitary matrix:
\begin{equation}U_\mathrm{bs} =  
   \begin{pmatrix} 
     1 &  0 & 0 & 0 \\
     0 & 1 &   0 & 0 \\
     0 &  0 & \cos \phi & -\sin \phi\\
     0 &  0 & \sin \phi & \cos \phi
   \end{pmatrix},\label{eq:u-general-bs}
 \end{equation} which can be interpreted as a beamsplitter matrix with transmisivity $t=\cos^2(\phi)$ and reflectivity $r=\sin^2(\phi)$. It should be noted that this circuit couples the 2--qubit state $\ket{10}\equiv\ket{2}_3$ to the discarded 2-qubit state $\ket{11}\equiv\ket{\ell}$, with a probability equal to $r$.

\section{Basis of operators and MUBs}\label{ap:bases}

To expand each of the channels $\mathcal{E}$ studied in this work we have chosen, as bases of unitary operators, the well known Sylvester's bases \cite{Sylvester, Singh2018}, which for any dimension $D$ can be written as:
\begin{equation}
  E_{n}\equiv E_{k l} = \sum_{m=0}^{D-1} \omega^{ml} \ketbra{m \oplus k}{m},
\end{equation}
where $k,l = 0,\dots, D-1$, $\omega = \exp(2 \pi i/D)$ is a root of unity and $\oplus$ is the  addition modulo--$D$. 

For the case $D=D_1=2$, the four operators are the Pauli operators together with the identity operator
\begin{equation} 
E_{00} = \mathbb{I}_2 \: \: , \: \: E_{01} = \sigma_z \: \: , \: \: E_{10} = \sigma_x \: \: , \: \: E_{11} = i \sigma_y \, ,
\end{equation}
from where we can obtain three abelian sets of two elements each: $\{ E_{00} , E_{01} \}$, $\{ E_{00} , E_{10} \}$ and $\{ E_{00} , E_{11} \}$. The three bases that diagonalize each of these sets, formed by eigenvectors of the Pauli operators, not only give a complete set of MUBs for $D_1$ (and, hence, an appropriate $2$--design for $\mathcal{H}_1$) but also have the property that the action of any of the four operators, $E_{kl}$, over any of the elements in the $2$--design, gives another element within the same MUB basis, except for a global phase. In fact, if $\ket{\psi_M^J}$ is one of the $D$ elements within the $J$-MUB, the following property is verified:  
\begin{equation}
E_{kl} \ket{\psi_M^J} = e^{i\alpha(k,l,M,J)} \ket{\psi_{M'}^J}.
\label{Op_on_MUBs}
\end{equation}

Analogously, in the case that $D=D_2=3$, we can obtain a $2$--design by extracting four abelian subsets from the nine operators $E_{kl}$. The first of them, $\{ E_{00},E_{01},E_{02} \}$, is diagonalized by the canonical basis
\begin{equation}
\mathcal{B}_0 = \left\{ \ket{0} , \ket{1} , \ket{2}   \right\} \equiv \left\{ (1,0,0) , (0,1,0) , (0,0,1)   \right\} \, .
\end{equation}
The next set, $\{ E_{00}, E_{10} , E_{20} \}$ is diagonalized by:
\begin{equation}
\mathcal{B}_1 = \left\{  \frac{(1,1,1)}{\sqrt{3}}  ,  \frac{(1,\omega,\omega^2)}{\sqrt{3}} , \frac{(1,\omega^2,\omega)}{\sqrt{3}}  \right\} \, ,
\end{equation}
where $\omega = \exp \left(2i\pi/3 \right)$, $\omega^2 = \omega^*$ and  $\omega^3 = 1$. It is clear that $\mathcal{B}_0$ and $\mathcal{B}_1$ are mutually unbiased. Moreover, by choosing $\mathcal{B}_2$ and $\mathcal{B}_3$ as the bases that diagonalize the sets $\{ E_{00} , E_{11} , E_{22} \}$ and $\{ E_{00} , E_{12} , E_{21} \}$, respectively, we get four MUBs in $D=3$ and hence a $2$--design in the corresponding Hilbert space, $\mathcal{H}_2$. 
Again, it is easy to check that the property given by Eq.~(\ref{Op_on_MUBs}) holds for the $9$ operators $E_{kl}$.

\section{Kraus decomposition of the implemented processes }\label{ap:krauss-processes}

All the processes implemented in this work correspond to a Kraus decomposition with a single unitary operator ($\E(\rho)=A\rho A^{\dag}$). We start by listing those of dimension $d=3$:
\begin{itemize}
   
  \item \texttt{H01}: $A \;=   \frac{1}{\sqrt{2}} \;(\ketbra{0}{0} - \ketbra{1}{1}+ \sqrt{2}\ketbra{2}{2}+\;\ketbra{0}{1} + \ketbra{1}{0} )$.
\item \texttt{H01 + LOSS 50\%}: $A \;=   \frac{1}{\sqrt{2}} \;(\ketbra{0}{0} - \ketbra{1}{1}+ \ketbra{2}{2}+\;\ketbra{0}{1} + \ketbra{1}{0} )$

\item \texttt{H12}: $A \;=   \frac{1}{\sqrt{2}} \;(\sqrt{2}\ketbra{0}{0} + \ketbra{1}{1}- \ketbra{2}{2}+\;\ketbra{1}{2} + \ketbra{2}{1} )$.

\item \texttt{H12 + LOSS 50\%}: $A \;=   \frac{1}{\sqrt{2}} \;(\ketbra{0}{0} + \ketbra{1}{1}- \ketbra{2}{2}+\;\ketbra{1}{2} + \ketbra{2}{1})$
\end{itemize}

Bellow, we list the unitary operators corresponding to the processes in $d=6$: 
\begin{itemize}
\item \texttt{ID + LOSS 50\%}: $A  =   \ketbra{0}{0} + \ketbra{1}{1}  + \ketbra{3}{3} + \ketbra{4}{4}+  \frac{1}{\sqrt{2}} \left( \ketbra{2}{2} +  \ketbra{5}{5}\right).$ 
\item \texttt{PHASE}: $A= (\ketbra{0}{0} + \ketbra{3}{3}) + e^{i\pi/3} \; \left(\ketbra{1}{1} + \ketbra{4}{4}\right) 
 +  \left( \ketbra{2}{2} +  \ketbra{5}{5}\right)$.
\item \texttt{PHASE + LOSS 50\%}: $A  =  ( \ketbra{0}{0} + \ketbra{3}{3} )  +e^{i\pi/3} \; \left(\ketbra{1}{1} + \ketbra{4}{4}\right) +  \frac{1}{\sqrt{2}} \left( \ketbra{2}{2} +  \ketbra{5}{5}\right)$.
\item \texttt{SWAP25}: $A=( \ketbra{0}{0} + \ketbra{3}{3})  + \; \;\;e^{i\pi/3} \; \left(\ketbra{1}{1} + \ketbra{4}{4}\right) +   \left( \ketbra{2}{5} +  \ketbra{5}{2}\right).$
\item \texttt{SWAP25 + LOSS 50\%}: $A =  (\ketbra{0}{0} + \ketbra{3}{3})  +   \;\;e^{i\pi/3} \; \left(\ketbra{1}{1} + \ketbra{4}{4}\right)
  +  \frac{1}{\sqrt{2}} \left( \ketbra{2}{5} +  \ketbra{5}{2}\right)$
\end{itemize}

\bibliography{001_references,002_additional_references}

\end{document}